\documentclass[rnote]{aa}

\usepackage{color}

\usepackage{graphicx}
\usepackage{txfonts}
\usepackage{natbib}

\newcommand{\oldbtxt}[1]{{#1}}

\bibpunct{(}{)}{;}{a}{}{,} 
\begin{document}


\title{Non-linear behaviour of XTE J1550-564 during its 1998-1999 outburst, revealed by recurrence analysis}
\author{Petra Sukov\'a\inst{1}
\and Agnieszka Janiuk\inst{1}
}

\institute{Center for Theoretical Physics,
Polish Academy of Sciences, Al. Lotnikow 32/46, 02-668 Warsaw, Poland \\
\email{psukova@cft.edu.pl}
}

\date{Received ...; accepted ...}
  
\abstract
{We study the X-ray emission of the microquasar XTE J1550-564 and analyze the properties of its light curves using the recurrence analysis method. The indicators for non-linear dynamics of the accretion flow are found in the very high state and soft state of this source. The significance of deterministic variability depends on the energy band. We discuss the non-linear dynamics of the accretion flow in the context of the disc-corona geometry and propagating oscillations in the accretion flow.}

\keywords{black hole physics; accretion, accretion disks; chaos; X-rays: binaries}
\authorrunning{P. Sukov\'a et al.}
\titlerunning{Non-linear behaviour of XTE J1550-564}

\maketitle


\section{Introduction} \label{s:Introduction}

The microquasar XTE J1550-564 is a low-mass X-ray binary, with a companion main sequence star, and was discovered in its 1998 outburst by the All-Sky Monitor onboard the RXTE satellite \citep{1998IAUC.6938....1S}. The radio-jets that accompany its activity were also detected, similarly to many other black hole X-ray binaries \citep{2001_Hannikainen_Astro_Space_S}. The compact object's mass is estimated at 8-11 Solar masses and thus it is considered to be a black hole \citep{2002ApJ...568..845O}. The source exhibits all the canonical spectral states typically associated with LMXBs.

The source's spectral analysis performed for the RXTE data of XTE J1550-564, that spans the period between September 23 and October 6, 1998, place this object in the very high state (VHS) category \citep{2003MNRAS.342.1083G}. Further analysis performed for this source by \cite{2004MNRAS.353..980K} showed that the temperature of the accretion disk is much lower than expected for the VHS luminosity, and suggested that this should be interpreted as the signature of the disk truncation at the radius of about 30 $R_{\rm g}$.

The intrinsic luminosity of the source peaks above $10^{39}$ erg/s, 
and during both very high and soft states it remains well above $2-3 \times 10^{38}$ erg/s. Therefore the derived accretion rate can be in this source equal to at least 25\% of the Eddington limit, and reaching even 
$0.8 \dot M_{\rm Edd}$ at the outburst peak.
In the standard accretion disk theory, such high values of $\dot M$
lead to the intrinsic thermal and viscous instabilities caused by the dominant radiation pressure. Such instabilities manifest themselves in the non-lienar
evolution of the disk temperature and density at very short timescales, and
consequently imprint a characteristic, time-dependent pattern in the emitted X-ray luminosity.

In this work, we study the properties of the XTE J1550-564 light curves. We implement our method, recently described in \cite{nas_nonlin}, and we study the light curves in the X-rays as taken by the RXTE satellite during the fall 1998 and spring 1999. The capabilities of this method were tested numerically in \cite{chaos_proc} and its dependence on parameters was discussed in \cite{Serbia_proc}. Here we focus on the spectral and temporal evolution of the microquasars's properties and its possible non-linear behaviour.
We discuss the results in the context of the thermal-viscous instability of a Keplerian accretion disc, as well as coupling of its variability with the behaviour of hard X-ray emitting, quasi-spherical corona.

\section{Observational data} \label{s:Data}

The overall light curve and hardness ratio of the outburst  of XTE J1550-564 shows two separated stages of high luminosity \citep{2000astro.ph..1460B}. The first one lasted from the beginning of September to the beginning of October 1998 (MJD 51063 to 51120), and the second one spanned from the mid of December 1998 to the mid of April 1999 (MJD 51160 to 51280).
From the list of the publicly available observations made by RXTE in NASA's HEASARC arxiv\footnote{\tt heasarc.gsfc.nasa.gov} we selected several observations, which cover the two epochs quite evenly and which are long enough for our purpose. The data sample is summarised in Table~\ref{Table_big}.

\section{Recurrence analysis and non-linear dynamics indicator} \label{s:method}
 
We study the properties of the light curves obtained in different energy bands by means of the time series analysis. Our aim is to find out if the time variability of the flux contains the information about the nature of the dynamical system behind the emission of the radiation. Mainly, we want to discover if the underlying system behaves according to deterministic non-linear dynamics, which is governed by a low number of equations rather than by stochastic variations. In the former case, the traces of the non-linear behaviour are imprinted into the outgoing radiation. 

 \begin{table*}[tb]
 \caption{
Non-linear dynamics indicator $\bar{\mathcal{S}}$ obtained for $N_\epsilon$ different values of recurrence threshold $\epsilon$ in different energy bands for three observations 30188-06-05-00 (MJD 51068), 30191-01-12-00 (MJD 51082) and 30191-01-33-00 (MJD 51108). The number of points used is $N=32768$ 
 } 
 \label{Table_bands}
 \centering
 \begin{tabular}{ccc|ccc|ccc|ccc}
\hline \hline

\multicolumn{3}{c}{MJD}&  \multicolumn{3}{c}{51068}&\multicolumn{3}{c}{51082}&\multicolumn{3}{c}{51108}\\
  \hline \hline

 & energy [keV] & channels & $\bar{\mathcal{S}}$ & $\sigma$ & $N_\epsilon$ & $\bar{\mathcal{S}}$ & $\sigma$ & $N_\epsilon$ & $\bar{\mathcal{S}}$ & $\sigma$ & $N_\epsilon$ \\

\hline
A&< 2.99&0 -- 7&0.45 & 0.22 & 6 & 0.61 & 0.39 & 6 & -0.27 & 0.46 & 8 \\
B&2.99 -- 5.12&8 -- 13& 1.77 & 0.36 & 8 & 2.58 & 0.42 &8 & 0.30 & 0.26 & 8 \\
C&5.12 -- 7.96&14 -- 21& 4.58 & 0.68 & 9& 4.38 & 0.63 &8 & 2.21 & 0.77 & 8\\
D&7.96 -- 12.99&21 -- 35& 6.94 & 1.13 & 8 &4.81 & 0.65 & 8 & 1.78 & 0.36 & 8\\
E&> 12.99& 36 -- 249 & 2.80 & 0.71 & 8 & - & - & - & - & - & -\\
\hline
F&< 12.99& 0 -- 35 & 6.69 & 1.47 & 8& 6.48 & 0.73 &8 & 2.92 & 0.71 &8\\

 \end{tabular}
 \end{table*}

In the preceding paper \citep{nas_nonlin} we developed the method, which can reveal such traces hidden in the observational data.  The method includes performing recurrence analysis on the observational time series, which is in our case the light curve, and on the set of surrogate data series. The results are then compared
\oldbtxt{according to discriminating statistics. Our null hypothesis is that the data are a product of a linearly autocorrelated Gaussian process, so that the data point is given by a linear combination of the preceding points and a contribution of uncorrelated Gaussian noise, $\xi(n)$, hence $x_n = a_0 + \sum_{k=1}^q a_k x_{n-k} +\xi(n)$. The properties of such time series are fully described by their Fourier spectrum.}
 The surrogate data are constructed in such a way that they share the value distribution and the Fourier spectrum with the original series \citep{Theiler199277,Schreiber2000346}. However, if the original time series is a product of a non-linear system, the dynamical attributes obtained by the analysis will differ from the surrogate series.
 
Here we use this method to study the outburst of XTE J1550-564, so we only briefly describe its main points. We normalise the light curve to have zero mean and unit variance, then we construct a bunch of 100 surrogates using the procedure {\tt surrogates} from TISEAN\footnote{\tt http://www.mpipks-dresden.mpg.de/\textasciitilde tisean/}\citep{1999chao.dyn.10005H}. 
\oldbtxt{ The recurrence analysis is performed on the basis of the software package described by \cite{Marwan2007237,RP-web}, which also yields  the cumulative histogram of diagonal lines. }
We compute the recurrence quantifiers for the data and the surrogates, using the educated guess of its parameters time delay $\Delta t$ and embedding dimension $m$ for a set of recurrence thresholds~$\epsilon$, such that the recurrence rate is approximately in the range (1\% -- 25 \%).

After computing the cumulative histograms of diagonal lines for the observed data and their surrogates, we estimate the R\'enyi's entropy $K_2$ by the linear regression of the cumulative histogram. We follow the rules given in \cite{nas_nonlin} Appendix~B, in particular that the minimal length of the diagonal lines must exceed the time delay and that there must be at least five points for the regression.

 As the discriminating statistic we choose the logarithm of the estimated R\'enyi's entropy, which we compute for each threshold for the observed time series $Q^{\rm obs}(\epsilon)=\ln (K_2^{\rm obs})$ and similarly for the set of surrogates $Q^{\rm surr}_i(\epsilon)$. We find the mean $\bar{Q}^{\rm surr}(\epsilon)$ and the standard deviation $\sigma_Q^{\rm surr}(\epsilon)$ of the set $Q^{\rm surr}_i(\epsilon)$.
Then we define the weighted significance as
\begin{equation}
\mathcal{S}(\epsilon) = \frac{N_{\rm sl}}{N^{\rm surr}} \mathcal{S}_{\rm sl} - {\rm sign}( Q^{\rm obs} (\epsilon) - \bar{Q}^{\rm surr}(\epsilon) ) \frac{N_{\mathcal{S}_K}}{N^{\rm surr}}  \mathcal{S}_{K_2}(\epsilon) , \label{significance}
\end{equation}
where $N_{\rm sl}$ is the number of surrogates, which have only short diagonal lines, and $N^{\rm surr}$ is the total number of surrogates, $\mathcal{S}_{\rm sl} =3$ and  $\mathcal{S}_{K_2}$ is the significance computed only from the surrogates, which have enough long lines according to the relation
\begin{equation}
\mathcal{S}_{K_2} (\epsilon) = \frac{| Q^{\rm obs} (\epsilon) - \bar{Q}^{\rm surr}(\epsilon) |}{\sigma_{Q^{\rm surr}(\epsilon)}}.
\end{equation}

Finally, we define the non-linear dynamics (NLD) indicator, $\bar{\mathcal{S}}$, as the average of $\mathcal{S}(\epsilon)$ over the set of $N_\epsilon$ different thresholds. We say that the observation shows traces of non-linear behaviour if it exceeds the chosen threshold, $\bar{\mathcal{S}} > 1.5$.

The accuracy of the resulting NLD indicator is quite a complicated issue, because the observational data with their uncertainness undergo several subsequent statistical processes. However, we can treat the resulting values $\mathcal{S}(\epsilon)$ as a small set of $N_\epsilon$ measurements of  $\bar{\mathcal{S}}$. As such, the uncertainty of NLD indicator is given as $\sigma = R/(2\sqrt{N_\epsilon})$, where R is the spread of the data set, that is $R={\rm max}(\mathcal{S}(\epsilon)) - {\rm min}(\mathcal{S}(\epsilon))$.

\section{Results} \label{s:results}

\subsection{Energy dependence of NLD indicator} \label{s:energy}
\begin{figure*}
\includegraphics[width=\columnwidth]{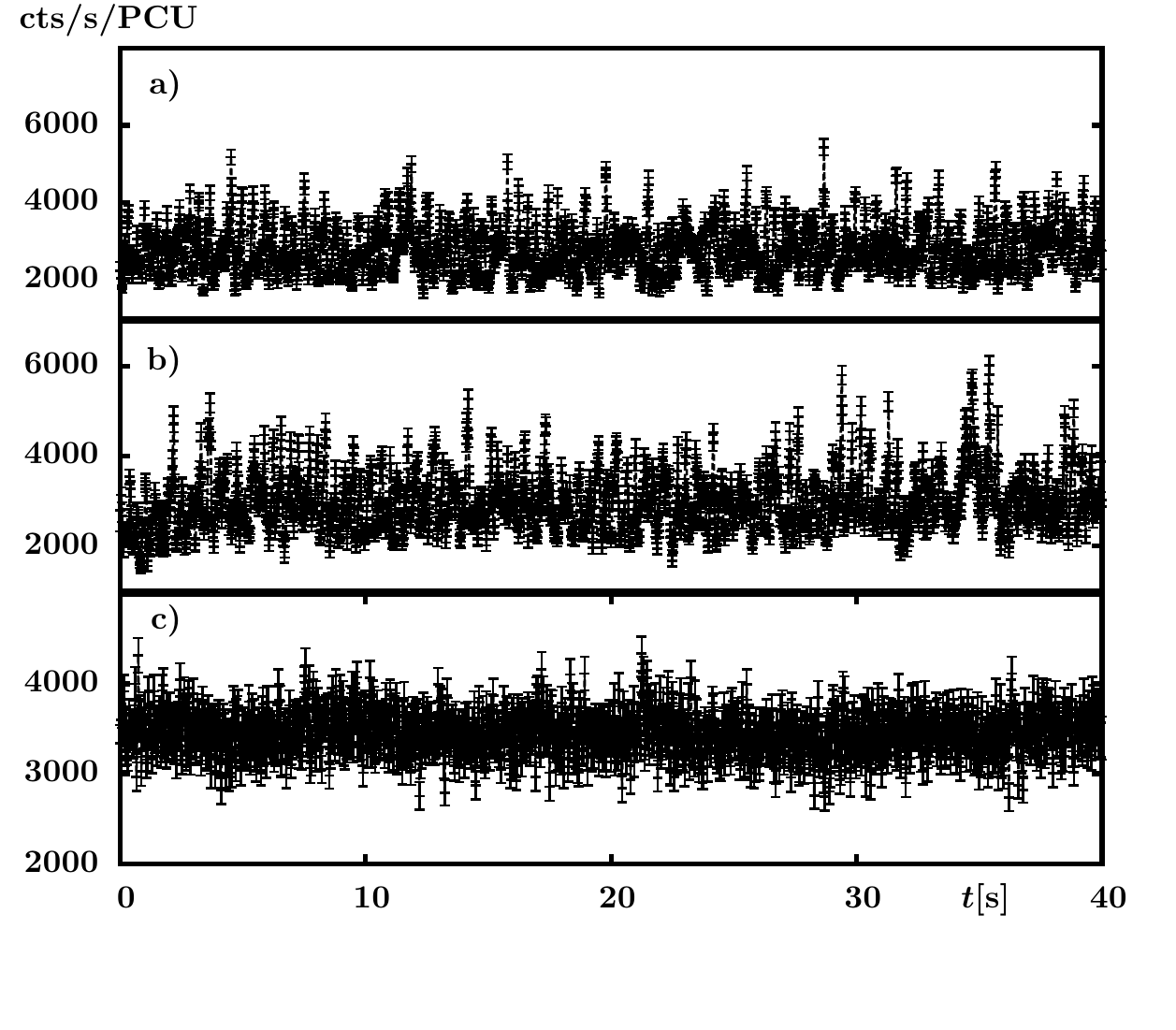}
\includegraphics[width=\columnwidth]{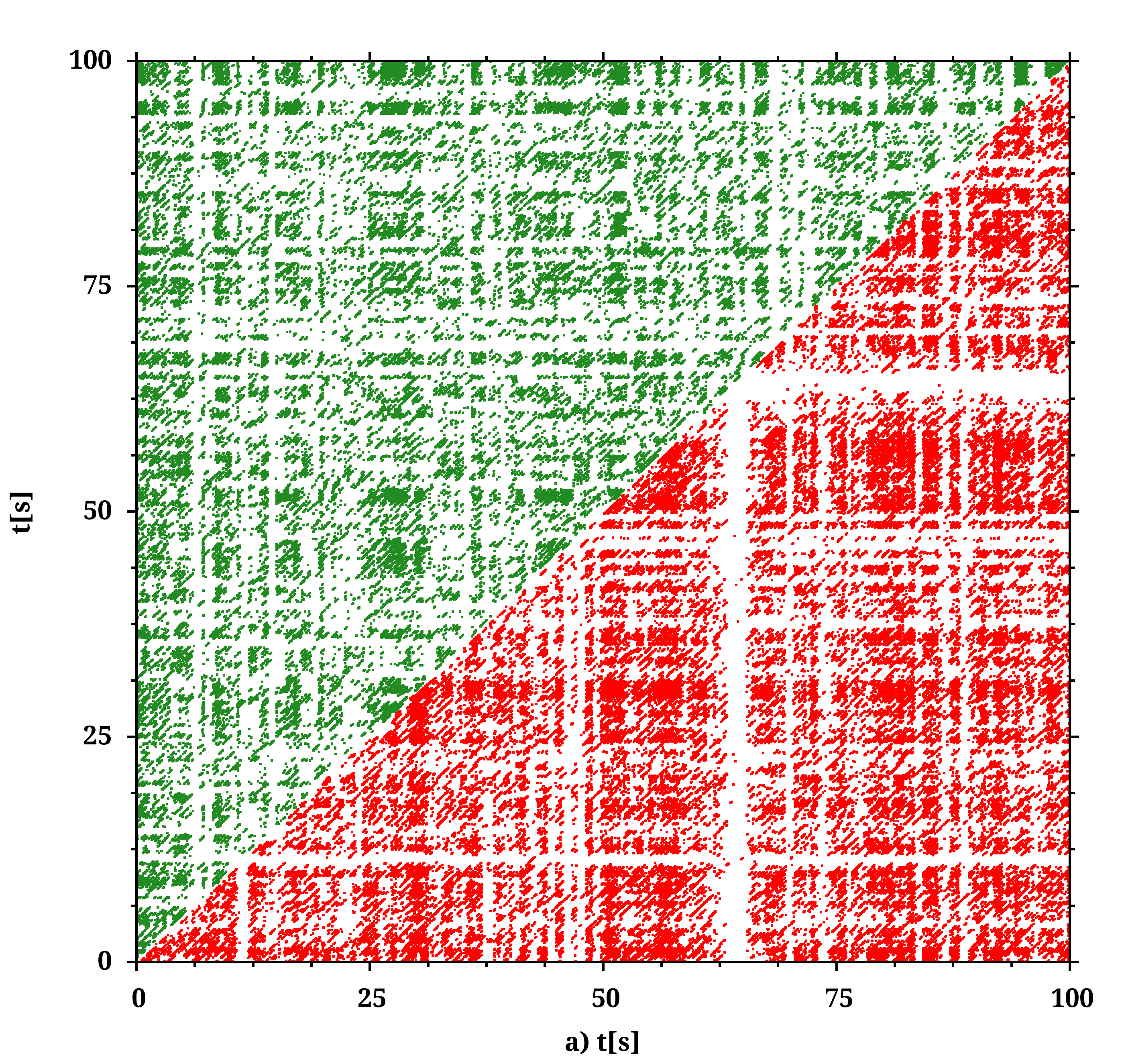}

\includegraphics[width=\columnwidth]{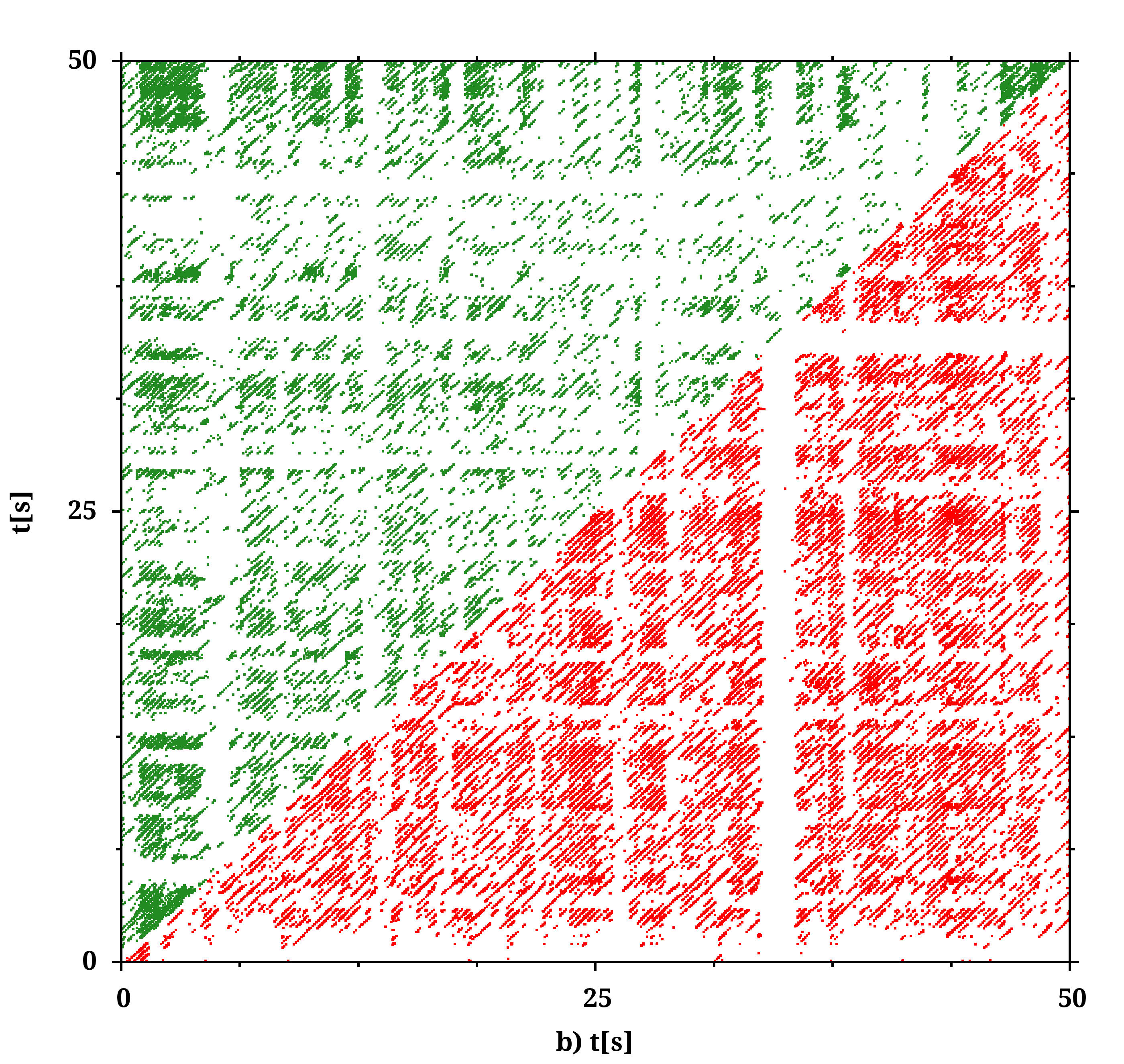}
\includegraphics[width=\columnwidth]{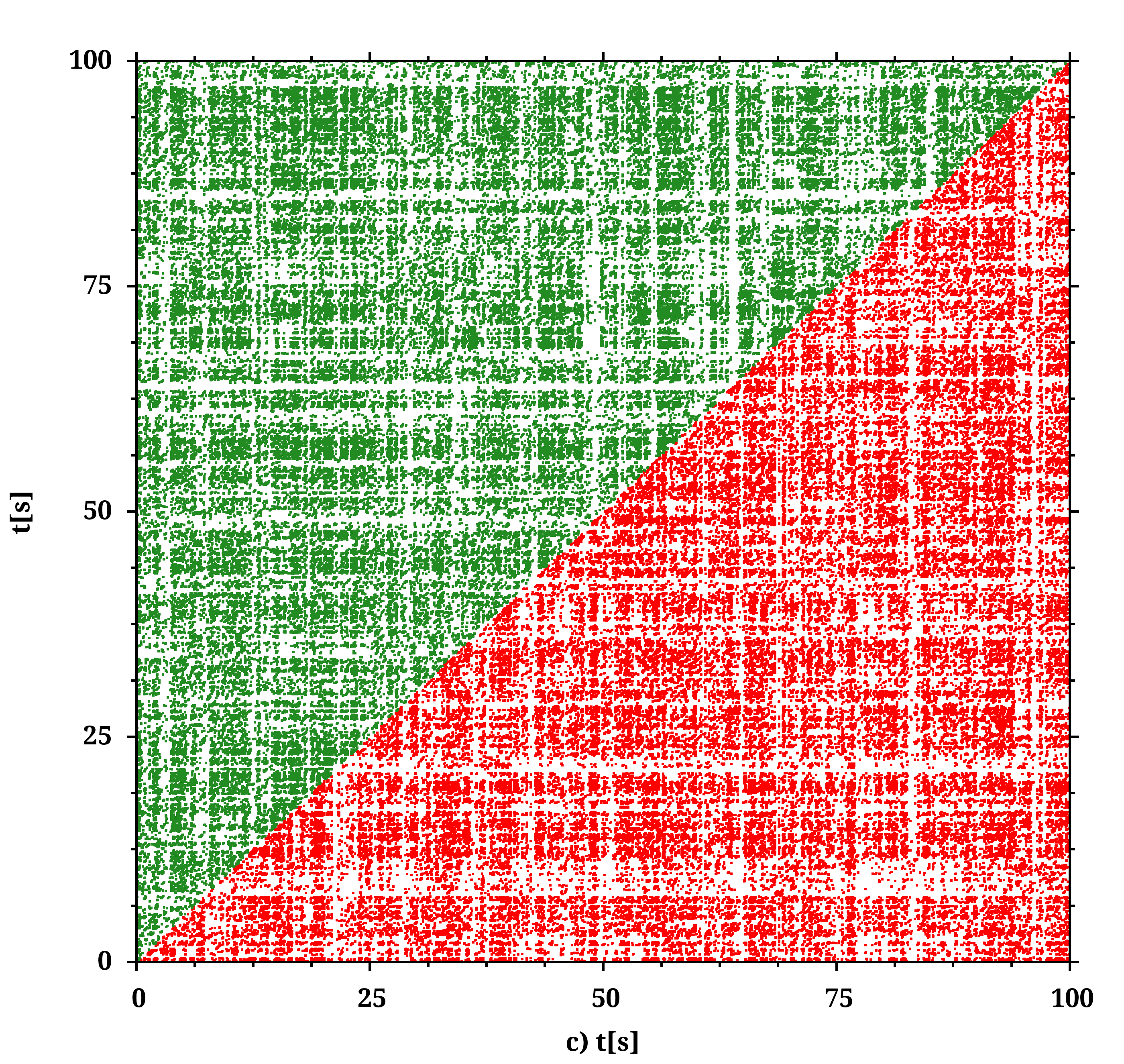}

\caption{\oldbtxt{Lightcurves in energy band F and recurrence plots for the observations (in red) and one surrogate (in green) for the three observations listed in Table~\ref{Table_bands}: 
a) MJD 51068 b) MJD 51082 c) MJD 51108.} }
\label{fig:lc}
\end{figure*}

We start our study with the observation 30188-06-05-00 taken on 12.9.1998 (MJD 51068). 
We extract the data with the time bin ${\rm d}t=0.025$s in six energy bands denoted as A -- F, which are listed in Table~\ref{Table_bands}. 
We take the longest available continuous part of the light curve up to $N_{\rm max}=32768$ points. We set the upper limit of the length of the studied time series in order to keep the numerical demands to a reasonable level. The light curve in band~F, the recurrence plot (RP)\footnote{Recurrence plot is a visualisation of the recurrence matrix, where the recurrence points are plotted as dots. Because the plot is symmetric with respect to the main diagonal, we plot only the triangle below/above the diagonal.} of the observation  and RP of one of the surrogates are plotted in Fig.~\ref{fig:lc}.

We compute the NLD indicator $\bar{\mathcal{S}}$ with the parameters $m=9$ and $k=4$. The highest $\bar{\mathcal{S}}$ is obtained for the energy band D, which covers the energy range $7.96 - 12.99$keV. 
On the other hand in the lowest energy band A, the NLD indicator is very small 
($\bar{\mathcal{S}}=0.45$). 

The indicator increases from band A to band D, slightly exceeding our threshold in band~B ($\bar{\mathcal{S}} = 1.77 > 1.5$) and giving quite a high value $\bar{\mathcal{S}} = 4.58$ in band C.
In the highest energies covered by band~E, the value decreases to $\bar{\mathcal{S}} = 2.80$.

\begin{figure}
\includegraphics[width=\columnwidth]{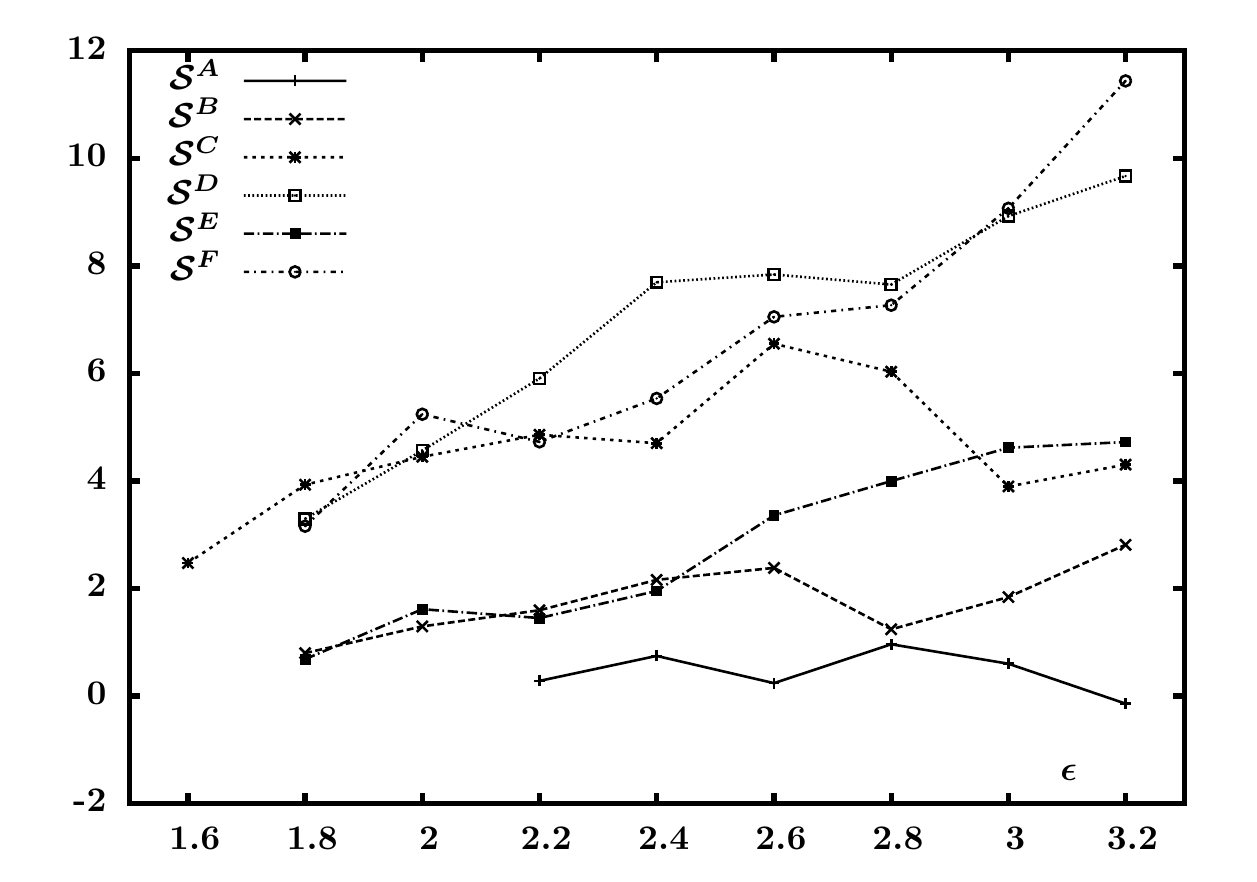}
\caption{\oldbtxt{Dependence of $\mathcal{S}$ on $\epsilon$ for the observation from MJD 51068 extracted in different energy bands.}}
\label{fig:sig_eps}
\end{figure}

We also provide results for the energy band F, which includes channels 0 -- 35 (< 12.99 keV). This band corresponds to the extraction of the data from the binned mode or the single-bit mode without energy filtering, so that such a light curve has the highest count rate. This band can also be compared with the results given in \cite{nas_nonlin}. The light curve in this band yields a high value $\bar{\mathcal{S}} = 6.69.$

We notice that the uncertainty $\sigma$ of NLD given in Table~\ref{Table_bands} in bands D and F is very high. The reason for this is that the significance $\mathcal{S}(\epsilon)$ is not independent on $\epsilon$, but rather grows with the threshold and the spread of the data set is high (see Fig.~\ref{fig:sig_eps}). 
This behaviour is not so prominent in any other case, and can be explained if the underlying cumulative histograms are inspected. 

The histograms of this observation for higher thresholds show an `elbow'
around $l \sim 3.5$s and the slope of the part after the 
elbow is about two times smaller than for the shorter lines. 
A similar elbow is not seen in the histograms of the surrogates, partly because they do not have many such long lines. 
For higher $\epsilon$, the elbow is more prominent in the observation's histogram, lowering the estimated value of $K_2^{\rm obs}$ and thus making the difference between the observation and the surrogates bigger. 

This feature can be explained as the presence of recurrence period in the system. The slope of the first part is higher, because it is more difficult to predict the evolution of the trajectory, when we have information about its piece shorter than the recurrence period (see \cite{Marwan2007237} for more details
about the properties of RPs).  Also a possible interplay between the value of the recurrence threshold and the noise in the data can play a role in this effect. 

Similar analysis of the energy dependence of NLD was also made for two later observations, 30191-01-12-00 taken on MJD 51082, and 30191-01-33-00 taken on MJD 51108 (denoted as b) and c) in Fig.~\ref{fig:lc}).

The second observation yields similar values of NLD in bands D and C and the overall band F gives the highest value. 
The third observation also has its highest significance in the band F, however for this observation the NLD indicator in band C is higher then in band D. 
The values in all bands are much less than for the first two observations, which suggests that the non-linear dynamics in the source fades away during the outburst.

\subsection{Temporal evolution of NLD indicator during the outburst} \label{s:temporal}

\begin{figure}
\includegraphics[width=\columnwidth]{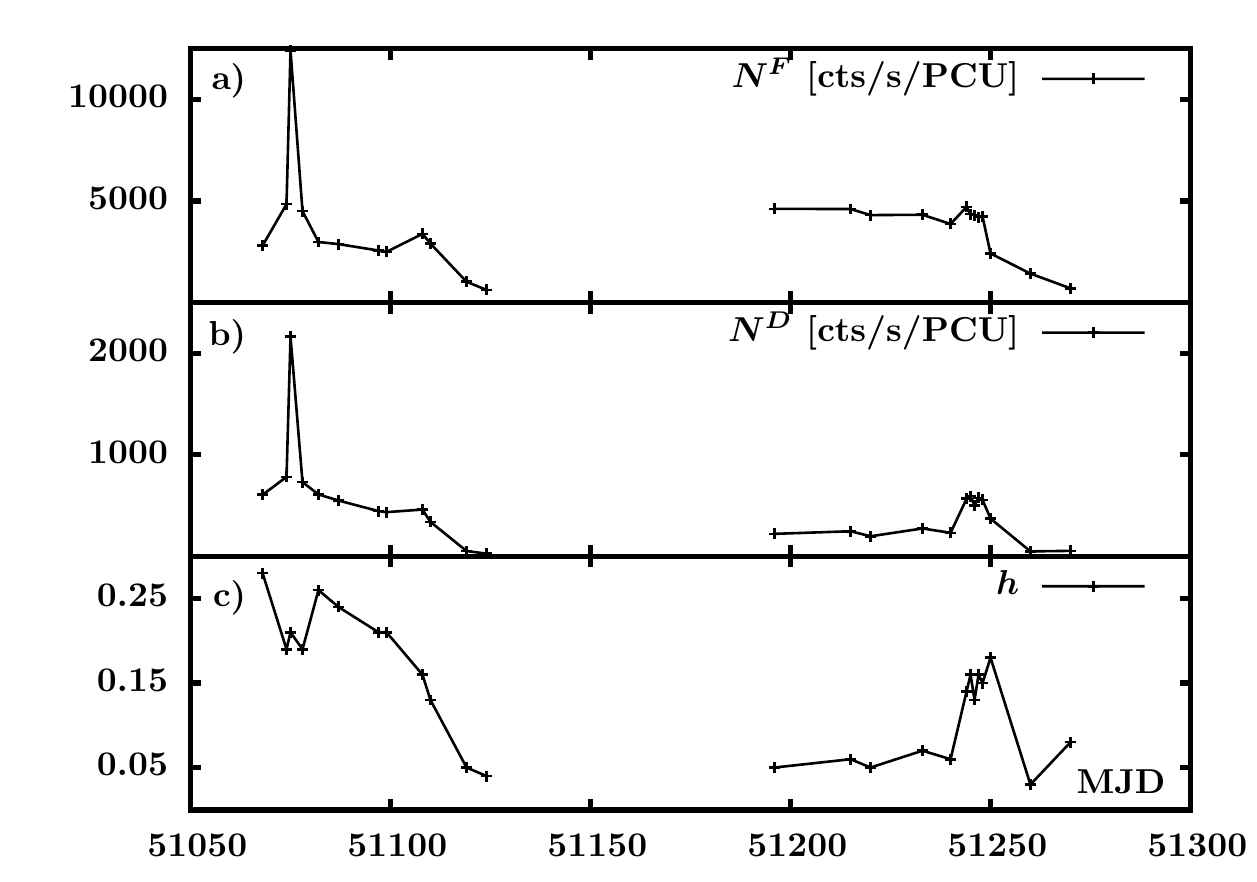}
\caption{Characteristics of the source XTE J1550-564 during both epochs of the outburst 1998/1999. 
a) Count rate in energy band F. b) Count rate in energy band D. c) Hardness ratio $h$.}
\label{fig:cts}
\end{figure}

We continue with the study of the temporal evolution of the NLD indicator during the outburst of XTE J1550-564. 
We extract the light curves in energy bands D and F for each observation from Table \ref{Table_big}. 
In order to use our procedure meaningfully, we need to have a sufficiently high number of counts in each time bin. Hence, the total count rate combined from all working proportional counter units (PCUs) in each energy band is an important quantity for setting the time bin ${\rm d}t$. 
We provide the values 
and the number of working PCUs 
in Table~\ref{Table_big}. 
The normalised count rate per PCU is given in Figs~\ref{fig:cts}a) and~\ref{fig:cts}b), respectively. 

We quantify the hardness ratio as the count rate in band D divided by count rate in lower energies,         $h = N_{\rm D} / (N_{\rm F} - N_{\rm D})$, which is given in 
Table~\ref{Table_big} and in Fig~\ref{fig:cts}c). 
Strong spectral transition from hard state, with $h \gtrsim 0.2$, to soft state, with $h \approx 0.05$, can be seen during the first stage  of the outburst. 
Hence, the count rate in band D drops down significantly in the later epoch. 
Therefore we do not compute the significance in band D for observations after 1.11.1998 (MJD 51119).

During the second stage of the outburst, the hardness ratio $h$ is much lower, with the exception of short time period around MJD 51250 (see \citet{2004MNRAS.353..980K}), when it reaches $h=0.18$. We map this time period more closely, choosing several observations between MJD 51240 and MJD 51250. Apart from those observations, the count rate in band D is low, so that applying our method on the light curve in band D is not possible due to the lack of counts in a short time bin. Because of this, we provide NLD indicator only for band F.

We use the time bin ${\rm d}t^F=0.032$s for light curves in band F and ${\rm d}t^D=0.025$s for light curves in band D. We find the educated guess of the embedding dimension $m$ and the time delay $\Delta t = k {\rm d}t$ as in \cite{nas_nonlin}. The values vary in the range $m \in [7,9]$ and $k \in [2,9]$. 
For each observation and each energy band, $N_\epsilon$ indicates the number of different $\epsilon$, for which the NLD indicator is obtained by averaging the significance.

\begin{figure}
\includegraphics[width=\columnwidth]{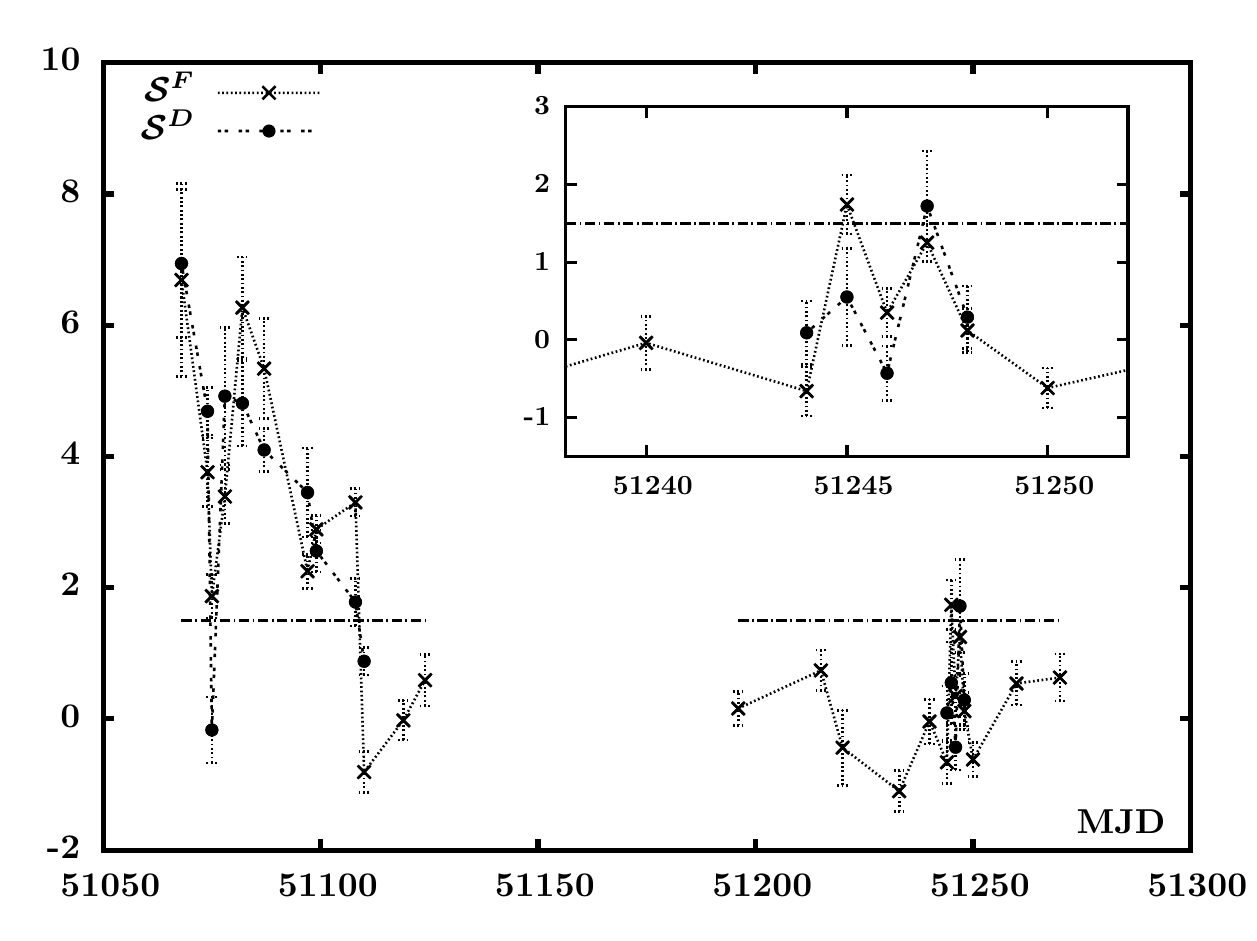}
\caption{NLD indicator in band D and F during the outburst of XTE J1550-564. 
The threshold 1.5 for significant result is indicated by the dashed horizontal line. }
\label{fig:NLD}
\end{figure}

Our results, which are summarised in 
Table~\ref{Table_big} and plotted in Fig.~\ref{fig:NLD}, show that the NLD indicator is higher in band D than in band F at the beginning of the outburst. This is in agreement with the result obtained for the observation 30188-06-05-00 (see Table~\ref{Table_bands}). 
The traces of non-linear dynamics are most prominent in the light curves consisting of counts in the energy range of 8-13 keV. Hence, the non-linear behaviour of the accreting system is expected to manifest mainly in the regions that emit radiation in energies 8-13 keV.

Later, this tendency is smeared and $\bar{\mathcal{S}}^F > \bar{\mathcal{S}}^D$ in some cases. The non-linear behaviour shifts into lower energies as the outburst continues, while its total importance decreases. This supports the trend, which we presented in section \ref{s:energy} for the three selected observations.

The obvious exception of the NLD indicator behaviour is the observation 30191-01-02-00, taken on MJD 51075, for which the NLD indicator in band F is slightly above the threshold, but the NLD indicator in band D is non-significant. 
This observation is special because it covers the peak of the outburst, during which the count rate is three times higher than in any other observation. Remarkably, the hardness ratio does not vary much, having very similar value, $h=0.21$, like for the other observations.

 \begin{table*}
 \small
 \caption{Observations selected for the analysis. Total count rate in energy bands F and D, the number of working PCUs and the hardness ratio $h = N_{\rm D} / (N_{\rm F} - N_{\rm D})$ are shown, along with  non-linear dynamics indicator $\bar{\mathcal{S}}$ in energy bands F and D, obtained for $N_\epsilon$ different values of recurrence threshold $\epsilon$ and the  time bin ${\rm d}t$, \oldbtxt{with uncertainty $\sigma.$}} 
  \label{Table_big}
 \begin{tabular}{ccc|ccc|c|cccc|cccc}
 \hline \hline
MJD & date & obsID & $N^F$ & $N^D$ & PCU & h & ${\rm d}t^F $ & $N_\epsilon^F$ & $\bar{\mathcal{S}}^F$ & $\sigma$ & ${\rm d}t^D $ & $N_\epsilon^D$ & $\bar{\mathcal{S}}^D$ & $\sigma$ \\
&&&[cnt/s]&[cnt/s]&num&&[s]&&&&[s]&&\\
\hline
51068 & 1998-09-12 & 30188-06-05-00 & 13991 & 3029 & 5 & 0.28 & 0.025 & 8 & 6.69 & 1.47 & 0.025 & 8 & 6.94 & 1.13 \\
51074 & 1998-09-18 & 30191-01-01-00 & 24222 & 3910 & 5 & 0.19 & 0.032 & 6 & 3.76 & 0.52 & 0.025 & 6 & 4.69 & 0.36 \\
51075 & 1998-09-19 & 30191-01-02-00 & 62059 & 10820 & 5 & 0.21 & 0.032 & 4 & 1.87 & 0.33 & 0.025 & 6 & -0.17 & 0.50\\
51078 & 1998-09-22 & 30191-01-08-00 & 22485 & 3651 & 5 & 0.19 & 0.032 & 8 & 3.39 & 0.41 & 0.025 & 7 & 4.92 & 1.05 \\
51082 & 1998-09-26 & 30191-01-12-00 & 14897 & 3050 & 5 & 0.26 & 0.032 & 7 & 6.27 & 0.77 & 0.025 & 8 & 4.81 & 0.65 \\
51087 & 1998-10-01 & 30191-01-18-00 & 14376 & 2750 & 5 &  0.24 & 0.032 & 7 & 5.34 & 0.76 & 0.025 & 7 & 4.10 & 0.33 \\
51097 & 1998-10-11 & 30191-01-28-01 & 12787 & 2224 & 5 &  0.21 & 0.032 & 6 & 2.25 & 0.26 & 0.025 & 8 & 3.45 & 0.68 \\
51099 & 1998-10-13 & 30191-01-29-00 & 12455 & 2172 & 5 &  0.21 & 0.032 & 7 & 2.89 & 0.21 & 0.025 & 7 & 2.56 & 0.32 \\
51108 & 1998-10-22 & 30191-01-33-00 & 16886 & 2304 & 5 &  0.16 & 0.032 & 7 & 3.30 & 0.21 & 0.025 & 8 & 1.78 & 0.36 \\
51110 & 1998-10-24 & 30191-01-34-00 & 14492 & 1693 & 5 &  0.13 & 0.032 & 4 & -0.81 & 0.31 & 0.025 & 6 & 0.88 & 0.21\\
51119 & 1998-11-01 & 30191-01-38-00 & 5078 & 258 & 5 & 0.05 & 0.032 & 4 & -0.02 & 0.30 & - & - & - \\ 
51124 & 1998-11-07 & 30191-01-40-00 & 3062 & 131 & 5 & 0.04 & 0.125 \tablefootmark{a}& 4 & 0.59 & 0.39 & - & - & - \\ \hline
51196 & 1999-01-18 & 40401-01-15-00 & 18438 & 886 & 4 & 0.05 & 0.032 & 4 & 0.16 & 0.26 & - & - & - \\
51215 & 1999-02-06 & 40401-01-31-00 & 23020 & 1235 & 5 & 0.06 & 0.032 & 8 & 0.74 & 0.31 & - & - & - \\
51220 & 1999-02-11 & 40401-01-35-00 & 12912 & 588 & 3 & 0.05 & 0.032 & 8 & -0.44 & 0.57 & - & - & - \\
51233 & 1999-02-24 & 40401-01-45-00 & 12950 & 822 & 3 & 0.07 & 0.032 & 8 & -1.10 & 0.31 & - & - & - \\
51240 & 1999-03-03 & 40401-01-49-00 & 19345 & 1151 & 5 & 0.06 & 0.032 & 6 & -0.04 & 0.34 & - & - & - \\
51244 & 1999-03-07 & 40401-01-52-00 & 14117 & 1701 & 3 &  0.14 & 0.032\tablefootmark{b} & 7 & -0.66 & 0.32 & 0.025\tablefootmark{c} & 8  & 0.09 & 0.41\\
51245 & 1999-03-08 & 40401-01-53-00 & 13076 & 1775 & 3 & 0.16 & 0.032 & 7 & 1.74 & 0.38 & 0.025 & 7  & 0.55 & 0.62 \\
51246 & 1999-03-09 & 40401-01-54-00 & 17210 & 1998 & 4 & 0.13 & 0.032 & 7 & 0.35 & 0.31 & 0.025 & 7  & -0.43 & 0.35\\
51247 & 1999-03-10 & 40401-01-55-00 & 16815 & 2301 & 4 & 0.16 & 0.032 & 7 & 1.25 & 0.24 & 0.025 & 7  & 1.72 & 0.71\\
51248 & 1999-03-11 & 40401-01-51-01 & 12699 & 1668 & 3 & 0.15 & 0.032 & 7 & 0.12 & 0.28 & 0.025 & 7  & 0.29 & 0.40\\
51250 & 1999-03-13 & 40401-01-57-00 & 4831 & 741 & 2 & 0.18 & 0.032 & 7 & -0.62 & 0.26 & - & - & - \\
51260 & 1999-03-23 & 40401-01-63-00 & 5687 & 191 & 4 & 0.03 & 0.032 & 8 & 0.54 & 0.33 & - & - & - \\
51270 & 1999-04-02 & 40401-01-71-00 & 3472 & 261 & 5 & 0.08 &0.032 & 10 & 0.63 & 0.36 & - & - & -\\ \hline
  \end{tabular}
 \tablefoot{ The recurrence analysis was done for a continuous part of the observation consisting of $N=32768$ points. The exceptions are the following:
 \tablefoottext{a}{$N=16507$},
 \tablefoottext{b}{$N=21437$},
 \tablefoottext{c}{$N=27439$}.
 }
 \end{table*}

On MJD 51110, the NLD indicator is less than one in both bands and the evidence for non-linear behaviour vanishes. 
The hardness ratio drops down to $h=0.13$, while the total count rate $N^F$ remains at similar level $\sim 14 000$ cts/s.
From that moment, not only the hardness ratio decreases further to $h\sim 0.05$, but the total count rate also declines significantly. 
NLD indicator shows no signs of non-linear dynamics. 

 After the source stays at a low luminosity level for about two months, the second stage of the outburst occurs. 
 This epoch spans approximately from the middle of January to the beginning of April 1999. 

During this epoch, the count rate in band F reaches similar values to the first stage ($\sim 15000$ cts/s) except for the outburst peak. 
Almost all observations from this stage including those with higher $h$ provide non-significant values of NLD indicator, with only slight glimpses of non-linear motion on MJD 51245 ($\bar{\mathcal{S}}^F=1.74$) and MJD 51247 ($\bar{\mathcal{S}}^D=1.72$).
Hence, there are only very weak signs of non-linear dynamics during the second epoch of the outburst and the behaviour of the source is considerably different from the first stage.

\section{Discussion} \label{s:discussion}

During the two epochs of the outburst the source XTE J1550-564 undergoes a complicated evolution of its 
total luminosity and the spectral and timing properties. Existence and properties of quasi-periodic oscillations (QPOs) were studied, for example, by \cite{2002ApJ...564..962R,2000ApJ...531..537S,2011MNRAS.411L..66H}. Detailed spectral study was provided by \cite{2000ApJ...544..993S}.

\cite{2004MNRAS.353..980K} discussed the spectral fitting of the data with respect to the geometry of the accretion flow.
They focus on the first epoch of the outburst, when the evolution goes from a low/hard state through a strong very high state, VHS and a weak VHS to the standard high/soft state. 
Authors show that the data are hardly compatible with the constant inner radius of the accretion disc. 
They propose a scenario of the evolution of the accretion flow during the initial state of the outburst, which involves the accretion disc truncated at higher radius than the ISCO, which is immersed in the hot Comptonizing corona. 
With the elapsed time, the disc propagates closer to the center, eventually reaching the ISCO around MJD 51105. 
Within next 12 days the remaining corona disappears and the source moves into the standard high/soft state.

The connection of the outburst time and spectral variability with the properties of companion star was shown by \cite{0004-637X-565-2-1161}. 
They proposed a scenario of the outburst, which begins with the accretion of low angular momentum gas released from a magnetic trap near the L1 point created by the companion star magnetic field. 
At the same time high angular momentum accretion also begins through the Roche lobe overflow forming an accretion disc further away. 
However, the low angular momentum accretion is much faster, operating near the free fall speed, and the electrons gains high energy. 
Close to the black hole, where a centrifugal barrier is met, a shock in this component could create and possibly oscillate, showing the non-linear behaviour and radiating in the highest energies.
This scenario explains the fast rise of the hard X-ray component versus much slower rise of the soft component, and it is compatible with the previously presented picture of the colder Keplerian disc slowly propagating inward through the already existing hot corona \citep{2004MNRAS.353..980K}.

This scenario is also viable in point of view of our results. 
We showed that at the beginning of the outburst the non-linear behaviour is the strongest in the higher energies, particularly in band D. 
We note that the drop down of the NLD indicator in the highest energy band E may be caused by a small count rate, and perhaps data from different instrument would be more suitable for the analysis in the highest energies. Later the non-linear dynamics is more prominent in lower energies (band C).

The oscillations of the accretion disc caused by the radiation pressure instability occurs typically on timescales of at least five to ten seconds. The exact value depends on the global parameters of the source, such as the black hole mass and viscosity in the accretion disk, and is triggered by the accretion rate being a large fraction of the Eddington. The threshold accretion rate for the instability depends weakly on the system parameters. The truncation of the inner disc, on the other hand, could shut down the instability in XTE J1550-564. For instance, the disc truncated at 30 Schwarzschild radii would be completely stable, if the accretion rate is below 0.7 of Eddington rate.  Therefore only the `memory' of the disc's oscillations would be kept in the hot corona, which is dynamically and radiatively coupled with the disc. 
It should also be noted, that instead of the regular outbursts on large amplitudes and long timescales, which are not observed in XTE J1550-564, a kind of flickering behaviour is possible, in case of a specific configuration of system parameters for which the disc is only marginally unstable (Grzedzielski et al., 2016, in prep.).

Here, we link the non-linear behaviour of the high energy emitting regions with the oscillations inside the corona, triggered by the formation and propagation of the accretion disc in the equatorial plane.
When the disc is more deeply immersed in the corona,
the non-linear behaviour shifts to the lower energies. 
When the disc reaches the ISCO,  and accretion rate is low, it stabilises and emits only thermal stochastic radiation. 
The oscillations of the corona cease and the corona itself is depleted within several days.

The propagation of oscillations through the corona can potentially lead to the evolution of the low frequency QPOs, modelled by \cite{Chakrabarti11042009} within an outburst scenario, that invoked formation of a shock in the inner region.
We can expect that during the flare peak (MJD 51075) any non-linear oscillatory behaviour is overpowered by some rapid, very energetic event accompanied by a strong stochastic emission, so that the NLD indicator is quite low.
This can be related to the jet launching because strong radio emission, which peaked two days after the X-rays, was observed preceded by a peak in optical band that occurred about one day after the X-ray peak \citep{0004-637X-565-2-1161}. A few days later the apparent superluminal motion of the ejecta in the radio band was observed \citep{2001_Hannikainen_Astro_Space_S}.

In the second epoch of the outburst, we detect no strong signs of non-linear behaviour, even in the short period between MJD 51244 and MJD 51250, when the hardness ratio increased significantly. 
Here, however, the overall evolution of the source is different. 
We first see the soft state with high luminosity and small $h$, compatible with an accretion disc spanned down to the ISCO, and then the increase of the high energy flux $N^D$ (and $h$).
Obviously, the mechanism of the corona creation in this case is different than in the first epoch and the non-linear mechanism leading to the oscillations is not triggered.

\section*{Acknowledgments}
This work was supported in part by the grant DEC-2012/05/E/ST9/03914 from the
Polish National Science Center.


\bibliographystyle{aa} 
\bibliography{Sukova} 


\end{document}